\documentclass[conference,twocolumn]{IEEEtran}

\usepackage{amsmath,epsfig,bm,amssymb,amsthm}
\usepackage{psfrag}
\usepackage{cite}
\ifCLASSINFOpdf
\else
\fi

\usepackage{amsmath,epsfig,bm,amssymb,amsthm,balance}

\newcommand{\dd}{{\mathrm d}}

\newcommand{\CN}{\mathcal{CN}}
\newcommand{\Et}{\mathrm{E}}
\newcommand{\bt}{\mathrm{b}}
\newcommand{\Var}{\mathrm{Var}}
\newcommand{\opt}{\mathrm{opt}}

\newcommand{\tzeta}{\tilde{\zeta}}

\newcommand{\sMRC}{\mathrm{semi-OPT1}}
\newcommand{\sOPT}{\mathrm{semi-OPT}}

\begin{document}
%
\title{Differential Amplify-and-Forward Relaying Using Linear Combining in Time-Varying Channels}

\author{
	\IEEEauthorblockN{M. R. Avendi\IEEEauthorrefmark{1}, Ha H. Nguyen\IEEEauthorrefmark{2} and Dac-Binh Ha\IEEEauthorrefmark{3}}
\IEEEauthorblockA{\IEEEauthorrefmark{1}University of California, Irvine, USA}
\IEEEauthorblockA{\IEEEauthorrefmark{2}University of Saskatchewan, Saskatoon, Canada}
\IEEEauthorblockA{\IEEEauthorrefmark{3}Duy Tan University, Danang, Vietnam}
m.avendi@uci.edu, ha.nguyen@usask.ca, hadacbinh@duytan.edu.vn}

\maketitle

\begin{abstract}
\label{abs}
Differential encoding and decoding can be employed to circumvent channel estimation in wireless relay networks. This article studies differential amplify-and-forward relaying using linear combining with arbitrary fixed combining weights in time-varying channels. An exact bit error rate (BER) analysis is obtained for this system using DBPSK modulation and over time-varying Rayleigh fading channels. The analysis is verified with simulation results for several sets of combining weights and in various fading scenarios.
\end{abstract}


\IEEEpeerreviewmaketitle

\section{Introduction}
\label{se:intro}


Differential amplify and forward (D-AF) relaying has been considered in \cite{DAF-MN-Himsoon,DAF-DDF-QZ,DAF-General} to circumvent channel estimation at both Relay and Destination. In D-AF system, information bits are differentially encoded at Source. The Relay's function is simply to multiply the received signals with a fixed amplification factor. At Destination, decision variables are computed for all links and then linearly combined. Computing the optimum combining weights requires the instantaneous CSI of Relay-Destination (RD) channels (unknown) and the amplification factor. Thus, the average fading powers of the RD channels have been used to define a set of fixed weights in \cite{DAF-MN-Himsoon,DAF-DDF-QZ,DAF-General}. For further reference, these weights are referred to as {semi-OPT1} weights. It was also pointed out that the performance of a D-AF system using the semi-OPT1 weights is difficult to derive \cite{DAF-MN-Himsoon,DAF-DDF-QZ,DAF-General}. Instead, the performance of a D-AF system, in \emph{slow-fading} environments, based on the {optimum} weights (assuming instantaneous CSI of RD links are available) was derived for benchmarking the performance of the system using the semi-OPT1 weights.

On the other hand, channels may vary in time due to mobility of users, and this would affect the system performance. In \cite{DAF-ITVT}, the authors analysed D-AF relaying using a new set of combining weights over time-varying Rayleigh fading channels. The new combining weights were computed based on the second-order statistics (variance and auto-correlation) of the transmission links and the amplification factor. For further reference, these weights are referred to as {semi-OPT2}. However, only a lower bound of the bit-error-rate (BER) was derived in \cite{DAF-ITVT}. When there is no access to the above information, Destination may simply use equal gain combining (EGC) (albeit with some performance penalty). Therefore, it is necessary to consider a D-AF relaying system using arbitrary fixed combining weights and analyse its performance.

Motivated by the above discussion, in this article, D-AF relaying using linear combining with \emph{arbitrary fixed combining weights} is analysed over \emph{time-varying} Rayleigh-fading channels. An exact average bit-error-rate (BER) expression of the system using DBPSK is obtained. Specifically, the analysis is verified with simulation results for semi-OPT1 and semi-OPT2 weights in various fading scenarios.

The outline of the paper is as follows. Section \ref{sec:system} describes the system model. In Section \ref{sec:combinig}, linear combining, differential detection and the performance of the system are considered. Simulation results are given in Section \ref{sec:sim}. Section \ref{sec:con} concludes the paper.

\emph{Notation}: $(\cdot)^*$, $|\cdot|$, $\Re\{\cdot\}$,$\Im\{\cdot\}$ denote conjugate, absolute value, the real part and the imaginary part of a complex number, respectively. $\mathcal{CN}(0,\sigma^2)$ stands for complex Gaussian distribution with mean zero and variance $\sigma^2$ and $\chi_2^2$ stands for chi-squared
distribution with two degrees of freedom. $\Et\{\cdot\}$ and $\Var\{\cdot\}$ are expectation and variance operations, respectively. Both $\exp(\cdot)$ and $e^{(\cdot)}$ indicate exponential function and $E_1(x)=\int \limits_{x}^{\infty} (e^{-t}/t)\dd t$ is the exponential integral function.

\section{System Model}
\label{sec:system}
\begin{figure}[t]
\psfrag {Source} [] [] [1.0] {Source}
\psfrag {Relay} [] [] [1.0] {Relay}
\psfrag {Destination} [] [] [1.0] {Destination}
\psfrag {h1} [] [] [1.0] {\;$h_1[k]$}
\psfrag {h2} [] [] [1.0] {\;\;$h_2[k]$}
\psfrag {h0} [] [] [1.0] {\;\;$h_0[k]$}
\centerline{\epsfig{figure={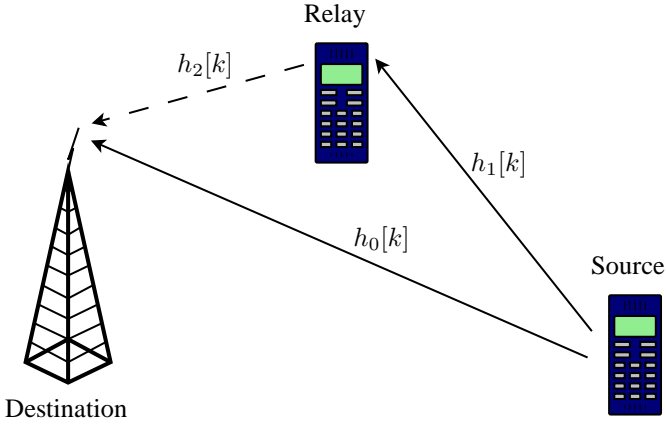},width=8.5cm}}
\caption{The wireless relay model under consideration.}
\label{fig:sysmodel}
\end{figure}

The wireless relay model under consideration is shown in Figure~\ref{fig:sysmodel}.
It has one source, one relay and one destination. Source communicates with Destination both directly and via Relay. Each node has a single antenna, and the communication between nodes is half duplex (i.e., each node is able to only send or receive in any given time). The channel coefficients at time $k$, from Source to Destination (SD), from Source to Relay (SR) and from Relay to Destination (RD) are shown with $h_0[k]$, $h_1[k]$ and $h_2[k]$, respectively. A Rayleigh flat-fading model is assumed for each channel, i.e., $h_i\sim \CN(0,\sigma_i^2), i=0,1,2$. The channels are spatially uncorrelated and changing continuously in time. It is assumed that all the channels follow a mobile-to-mobile channel model \cite{m2m-Akki}. 
Let, $f_i,\; i=0,1,2$ are the maximum normalized Doppler frequencies induced by the motion of Source, Relay and Destination in the SD, SR and RD links.
The time correlation between two channel coefficients, $n$ symbols apart, follows the Akki and Haber's model \cite{m2m-Akki}
\begin{gather}
\label{eq:phi0}
\varphi_i(n)=\Et\{h_i[k] h_i^*[k+n]\}=\sigma_i^2 J_0(2\pi f_i n) J_0(2\pi f_j n) 
\end{gather}
where $i=0,1,2,$ $j=2,0,1,$ $J_0(\cdot)$ is the zeroth-order Bessel function of the first kind. The normalized Doppler frequency is a function of the velocity of the nodes. A higher velocity leads to a higher Doppler value and hence a lower time-correlation between the channel coefficients. It should also be noted that the mobile-to-mobile channel model \cite{m2m-Akki} is a more general model which includes the fix-to-mobile model \cite{microwave-jake} as well. For instance, in case of fixed Destination 
$f_2=0$ and then  $\varphi_0(n),\varphi_2(n)$ change to the Jakes' fading model \cite{microwave-jake} defined with one Bessel function.

Let $\mathcal{V}=\{-1,+1\}$ be the set of BPSK symbols. Information bit at time $k$ is transformed to an BPSK symbol $x[k]\in \mathcal{V}$. Before transmission, the symbols are encoded differentially as

\begin{equation}
\label{eq:s-source}
s[k]=x[k] s[k-1],\quad s[0]=1.
\end{equation}


The transmission process is divided into two phases. 
Assume that, in phase I,  symbol $s[k]$ is transmitted by Source to Relay and Destination. Let $P_0$ be the average Source power per symbol. The received signals at Destination and Relay are

\begin{equation}
\label{eq:y0}
y_0[k]=\sqrt{P_0}h_0[k]s[k]+z_0[k],
\end{equation}
\begin{equation}
\label{eq:relay_rx}
y_1[k]=\sqrt{P_0}h_1[k]s[k]+z_1[k]
\end{equation}
where $z_0[k],z_1[k]\sim \CN(0,N_0)$ are noise components at Destination and Relay, respectively. It is easy to see that, for given $s[k]$, $y_0[k]\sim \CN(0,N_0(\rho_0+1)),$ where $\rho_0$ is the averaged received SNR per symbol from the direct path defined as
$
\label{eq:rho0}
\rho_0=\frac{P_0 \sigma_0^2}{N_0}.
$
Also, the average received SNR per symbol at Relay is defined as
$
\label{eq:rho_1}
\rho_1=\frac{P_0\sigma_1^2}{N_0}.
$
The received signal at Relay is then multiplied by an amplification factor $A$, and re-transmitted to Destination. 
Here, $A$ can be any arbitrarily fixed value.

The corresponding received signal at Destination is

\begin{equation}
\label{eq:y}
y_2[k]=A \; h_2[k]y_1[k]+z_2[k],
\end{equation}
where $z_2[k]\sim \mathcal{CN}(0,N_0)$ is the noise at Destination. 
After some manuipulations, it yields that
\begin{equation}
\label{eq:y2}
y_2[k]= A\; \sqrt{P_0}h[k]s[k]+z[k],
\end{equation}
where $h[k]=h_1[k]h_2[k]$ is the equivalent double-Rayleigh channel with zero mean and variance $\sigma_1^2 \sigma_2^2$ \cite{SPAF-P} and
$
\label{eq:z[k]}
z[k]=A\; h_2[k]z_1[k]+z_2[k]
$
is the equivalent noise. It should be noted that for a given $h_2[k]$, $z[k]$ is a complex Gaussian random variable with zero mean and variance
$
\label{eq:sig2wk}
\sigma_{z}^2=N_0(1+A^2 \; |h_2[k]|^2)
$
and hence $y_2[k]$, conditioned on $s[k]$ and $h_2[k]$ is a complex Gaussian random variable with zero mean and variance $(\rho_2+1)\sigma_{z}^2$. Here, $\rho_2$ is the average received SNR per symbol from the cascaded path at Destination, conditioned on $h_2[k]$. It is defined as
$
\label{eq:rhok}
\rho_2= \frac{A^2 \rho_1 |h_2[k]|^2}{1+A^2|h_2[k]|^2}.
$

\section{Detection and Performance Analysis}
\label{sec:combinig}

\subsection{Linear Combining and Differential Detection}
Given two consecutive received symbols at a time, non-coherent detection of the transmitted symbols can be obtained. Then, the decision variables for the direct and cascaded links are computed from the two latest symbols as
\begin{align}
\label{eq:zeta_0}
\zeta_0= \Re\{y_0^*[k-1] y_0[k] \}, \\
\label{eq:zeta_1}
\zeta_2= \Re\{ y_2^*[k-1] y_2[k] \}.
\end{align}
To achieve the cooperative diversity, the decision variables from the two phases are linearly combined as
\begin{equation}
\label{eq:zeta}
\zeta=w_0 \zeta_0+w_2 \zeta_2,
\end{equation}
where $w_0,w_2$ are the combining weights.

Finally, the output of the combiner is used to decode the transmitted signal as
\begin{equation}
\label{eq:ml-detection}
\hat{x}[k]= 
\begin{cases}
-1 & \mbox{if} \;\; \zeta<0\\
+1 & \mbox{if} \;\; \zeta>0.
\end{cases}
\end{equation}

In the next section, the performance of this system is analysed.

\subsection{Error Performance Analysis}
\label{sec:performance}

The transmitted symbols are equally probable and, without loss of generality, assume that symbol $x[k]=+1$ is transmitted and it is decoded erroneously as $\hat{x}[k]=-1$ by the decoder. The BER can be expressed as
\begin{multline}
\label{eq:PbE}
P_{\bt}(E)=\Pr(\zeta<0, x[k]=+1)
=\Pr(\tzeta_0+\tzeta_2<0)=\\
\int \limits_{-\infty}^{\infty} \int \limits_{-\infty}^{-\beta} f_{\tzeta_0}(\beta) f_{\tzeta_2}(r) \dd r \dd \beta
=\int \limits_{-\infty}^{\infty} f_{\tzeta_0}(\beta) F_{\tzeta_2}(-\beta)\dd \beta,
\end{multline}
where $\tzeta_0=w_0\zeta_0,\; \tzeta_2=w_2\zeta_2.$ Also, $f_{\tzeta_0}(\cdot)$ is the pdf of $\tzeta_0$ and $F_{\tzeta_2}(\cdot)$ is the cdf of $\tzeta_2$. To compute these function, first, $\zeta_0$ and $\zeta_2$ are simplified.

For time-varying channels, individual channels are expressed by an AR(1) model as
\begin{gather}
\label{eq:ARi}
h_i[k]=\alpha_i h_i[k-1]+\sqrt{1-\alpha_i^2} e_i[k],\quad i=0,1, 2
\end{gather}
where $\alpha_i=\varphi_i(1)/\sigma_i^2$ is the auto-correlation of the $i$th channel and $e_i[k]\sim \mathcal{CN}(0,\sigma_i^2)$ is independent of $h_i[k-1]$. Based on these expressions, a first-order time-series model has been derived in \cite{DAF-ITVT} to characterise the evolution of the cascaded channel in time. The time-series model of the cascaded channel is given as (the reader is referred to \cite{DAF-ITVT} for the detailed derivations/verification)
\begin{equation}
\label{eq:ARmodel}
h[k]=\alpha h[k-1]+\sqrt{1-\alpha^2}\ h_2[k-1]e_1[k],
\end{equation}
where $\alpha=\alpha_1 \alpha_2 \leq 1$ is the equivalent auto-correlation of the cascaded channel, which is equal to the product of the auto-correlations of individual channels, and $e_1[k]\sim \mathcal{CN}(0,\sigma_1^2)$ is independent of $h[k-1]$.
It should be noted that these time-series models are used only for performance analysis purpose and not for generating channel coefficients in the simulation.

Using \eqref{eq:ARi} and \eqref{eq:ARmodel} into $y_0[k],y_2[k]$ one has
\begin{equation}
\label{eq:y0k}
y_0[k]=\alpha_0 x[k] y_0[k-1]+\widetilde{z}_0[k],
\end{equation}
$
\label{eq:n0}
\widetilde{z}_0[k]=z_0[k]- \alpha_0 x[k] z_0[k-1]
+ \sqrt{1-\alpha_0^2} \sqrt{P_0} s[k]e_0[k].
$
\begin{equation}
\label{eq:yk}
y_2[k]=\alpha x[k] y_2[k-1]+\widetilde{z}[k],
\end{equation}
$
\label{eq:ni}
\widetilde{z}[k]=z[k]- \alpha x[k] z[k-1]
+ \sqrt{1-\alpha^2} A\sqrt{P_0}h_2[k-1]s[k]e_1[k].
$

Then by substituting \eqref{eq:y0k} and \eqref{eq:yk} into \eqref{eq:zeta_0} and \eqref{eq:zeta_1}, one has
\begin{align}
\label{eq:z0_simp1}
\zeta_0= \Re \left\lbrace \alpha_0 x[k] |y_0[k-1]|^2+y_0^*[k-1]\widetilde{z}_0[k] \right\rbrace \\
\label{eq:z1_simp1}
\zeta_2=\Re \left\lbrace \alpha x[k] |y_2[k-1]|^2+y_2^*[k-1]\widetilde{z}[k]\right\rbrace.
\end{align}
It is seen that, for given $y_0[k-1]$, $\zeta_0$ is a combination of complex Gaussian random variables with conditional mean and variance computed as 
\begin{equation}
\label{eq:mu_z0}
\mu_{\zeta_0}=\Et \{ \zeta_0 | y_0[k-1]\} 
=  \frac{\alpha_0  \rho_0}{\rho_0+1} |y_0[k-1]|^2,
\end{equation}
\begin{multline}
\label{eq:var_z0}
\Sigma_{\zeta_0}=\Var\{\zeta_0|y_0[k-1]\}
\\= \frac{1}{2}N_0 \left( 1+ \frac{\alpha_0^2\rho_0}{\rho_0+1} +
(1-\alpha_0^2) \rho_0
 \right)|y_0[k-1]|^2.
\end{multline}

Also, for given $y_2[k-1]$ and $h_2[k-1]$, $\zeta_2$ is a combination of complex Gaussian random variables and hence it is Gaussian as well with conditional mean and variance computed as 
\begin{equation}
\label{eq:mu_z1}
\mu_{\zeta_2}=\Et \{ \zeta_2 | y_2[k-1],h_2[k-1]\} 
= \frac{\alpha \rho_2}{\rho_2+1} |y_2[k-1]|^2,
\end{equation}
\begin{multline}
\label{eq:var_z1}
\Sigma_{\zeta_2}=\Var\{\zeta_2|y_2[k-1],h_2[k-1]\}
=\\ \frac{1}{2}\sigma_z^2 \left( 1+ \frac{\alpha^2\rho_2}{\rho_2+1} +
(1-\alpha^2) \rho_2 \right)|y_2[k-1]|^2.
\end{multline}
From now on, the time index $[k-1]$ is omitted for simplicity.

From \eqref{eq:mu_z0} and \eqref{eq:var_z0}, $\tzeta_0\sim \CN(w_0 \mu_{\zeta_0},w_0^2\Sigma_{\zeta_0})$ and its conditional pdf is written as
\begin{equation}
\label{eq:fz0_y0}
f_{\tzeta_0}(\beta|y_0)=\frac{1}{\sqrt{2\pi w_0^2 \Sigma_{\zeta_0}}} \exp \left( - \frac{(\beta-w_0\mu_{\zeta_0})^2}{2w_0^2\Sigma_{\zeta_0}}\right).
\end{equation}
Since $y_0\sim \CN(0,N_0(\rho_0+1))$, it follows that $|y_0|^2\sim 0.5{N_0(\rho_0+1)}\chi_2^2$. By taking the expectation of \eqref{eq:fz0_y0} over the distribution of $|y_0|^2$, the pdf of $\tzeta_0$ is obtained as 
\begin{equation}
\label{eq:f_z0}
f_{\tzeta_0}(\beta)=
\begin{cases}
\frac{1}{b_0}\exp\left(\frac{\beta}{c_0} \right) & , \;\; \beta\leq 0 \\
\frac{1}{b_0}\exp\left(\frac{\beta}{d_0} \right) & , \;\; \beta\geq 0
\end{cases}
\end{equation}
$b_0={w_0 N_0(1+\rho_0)}, 
c_0=0.5{w_0 N_0(1+(1-\alpha_0)\rho_0)}, 
d_0=-0.5{w_0 N_0(1+(1+\alpha_0)\rho_0)}.
$


On the other hand, from \eqref{eq:mu_z1} and \eqref{eq:var_z1}, $\tzeta_2\sim \CN(w_2 \mu_{\zeta_2},w_2^2\Sigma_{\zeta_2})$ and its conditional pdf is written as
\begin{equation}
\label{eq:fz1_y2_h2}
f_{\tzeta_2}(\beta|y_2,h_2)=\frac{1}{\sqrt{2\pi w_2^2\Sigma_{\zeta_2}}} \exp \left( - \frac{(\beta-w_2\mu_{\zeta_2})^2}{2w_2^2\Sigma_{\zeta_2}}\right).
\end{equation}
Since, conditioned on $h_2$, $y_2\sim \CN(0,\sigma_{z}^2(\rho_2+1))$, one concludes that $|y_2|^2\sim 0.5{\sigma_z^2 (\rho_2+1)}\chi_2^2$. By taking the expectation of \eqref{eq:fz1_y2_h2} over the distribution of $|y_2|^2$, one has 
\begin{equation}
\label{eq:f_z1_h2}
f_{\tzeta_2}(\beta|h_2)=
\begin{cases}
\frac{1}{b_2} \exp\left(\frac{\beta}{c_2} \right) & , \;\; \beta\leq 0 \\
\frac{1}{b_2} \exp\left(\frac{\beta}{d_2} \right) & , \;\; \beta\geq 0
\end{cases}
\end{equation}
$
b_2={w_2 \sigma_z^2(\rho_2+1)}, 
\label{eq:b2c2d2}
c_2=0.5{w_2 \sigma_z^2(1+(1-\alpha)\rho_2)}, 
d_2=-0.5{w_2 \sigma_z^2(1+(1+\alpha)\rho_2)},
$
are functions of random variable $\lambda=|h_2|^2$ whose pdf is $f_{\lambda}(\lambda)=(1/\sigma_2^2)\exp(\lambda/\sigma_2^2)$.

Thus, the cdf of $\zeta_2$ conditioned on $h_2$ is obtained as
\begin{equation}
\label{eq:F_z1_h2}
F_{\tzeta_2}(\beta|h_2)=
\begin{cases}
\frac{c_2}{b_2} \exp\left(\frac{\beta}{c_2} \right) & , \;\; \beta\leq 0\\
1+\frac{d_2}{b_2} \exp\left(\frac{\beta}{d_2}\right) & , \;\; \beta\geq 0.
\end{cases}
\end{equation}

Using \eqref{eq:f_z0} and \eqref{eq:F_z1_h2}, \eqref{eq:PbE} can be evaluated as
\begin{multline}
\label{eq:Pb1_sim1}
P_{\bt}(E|h_2)= \int \limits_{-\infty}^{\infty} f_{\tzeta_0}(\beta) F_{\tzeta_2}(-\beta|h_2) \dd \beta \\
=\int \limits_{-\infty}^{0} \frac{1}{b_0} e^{ \frac{\beta}{c_0}} \left(1+ \frac{d_2}{b_2} e^{- \frac{\beta}{d_2}} \right) \dd \beta
+\int \limits_{0}^{\infty} \frac{1}{b_0} e^{\frac{\beta}{d_0}} \left( \frac{c_2}{b_2} e^{- \frac{\beta}{c_2}} \right) \dd \beta \\
= \frac{c_0}{b_0}+\frac{c_0 d_2^2}{b_0b_2(d_2-c_0)}+\frac{d_0 c_2^2}{b_0b_2(d_0-c_2)}.
\end{multline}

By taking the final average over the distribution of $\lambda=|h_2|^2$, one has
$
\label{eq:Pb_sim2}
P_{\bt}(E)= I_1(\alpha_0)+I_2(\alpha_0,\alpha)+I_3(\alpha_0,\alpha).
$
The first term is obtained as
$
\label{eq:I1}
I_1(\alpha_0)=\frac{c_0}{b_0}= \frac{1+(1-\alpha_0)\rho_0}{2(1+\rho_0)}.
$
Interestingly, the term $I_1(\alpha_0)$, which corresponds to the direct link, is the same as the BER of DBPSK of point-to-point communications over fast-fading channels derived in \cite[eq.8.230]{DigComFad-Simon}.
Also, $I_2(\alpha_0,\alpha)$ and $I_3(\alpha_0,\alpha)$ are derived as
\begin{multline}
\label{eq:I2}
I_2(\alpha_0,\alpha)=\frac{c_0}{b_0} \int \limits_{0}^{\infty} \frac{d_2^2}{b_2(d_2-c_0)}f_{\lambda}(\lambda) \dd \lambda
\\=
\tilde{B}_3 \left(1+\frac{\tilde{B}_1-\tilde{B}_2}{\sigma_2^2}\exp\left(\frac{\tilde{B}_2}{\sigma_2^2}\right)E_1\left(\frac{\tilde{B}_2}{\sigma_2^2}\right) \right)
\end{multline}
\begin{figure*}
\begin{gather}
\label{eq:B1bB2bB3b}
\tilde{B}_2=\frac{(w_2(2+\alpha)+w_0+w_0(1-\alpha_0)\rho_0)\rho_1+w_0(1+(1-\alpha_0)\rho_0)+2w_2}{A^2w_2(1+\rho_1)(1+(1+\alpha)\rho_1)}
\end{gather}
\end{figure*}
\begin{equation}
\label{eq:I3}
I_3(\alpha_0,\alpha)=\frac{d_0}{b_0} \int \limits_{0}^{\infty} \frac{c_2^2}{b_2(d_0-c_2)}f_{\lambda}(\lambda) \dd \lambda
=-I_2(-\alpha_0,-\alpha).
\end{equation}
with $\tilde{B}_1=2/(A^2(1+(1+\alpha)\rho_1)),\tilde{B}_3=-(1+(1-\alpha_0)\rho_0)/(2\tilde{B}_1A^2(1+\rho_0)(1+\rho_1)),\tilde{B}_2$ defined on the top of the next page.

As can be seen, the obtained BER expression depends on the combining weights through $\tilde{B}_2$ and it gives the BER of the D-AF relaying system using DBPSK and arbitrary fixed combining weights in time-varying Rayleigh fading channels. 
Moreover, it is seen that the obtained BER expression depends on the channel auto-correlations through the defined parameters. This dependency on the channel auto-correlations leads to a performance degradation in fast-fading channels and eventually, at high transmit power, the BER reaches an error floor. The amount of error floor can be obtained as
$
\label{eq:PEf}
\lim \limits_{P_0/N_0 \rightarrow \infty} P_{\bt}(E), 
$
which is straightforward and not expressed here due to space limit.

\section{Numerical Results}
\label{sec:sim}
As mentioned before, the analysis in the previous section works for arbitrary fixed combining weights. In this section, the analysis is verified for two sets of combining weights, semi-OPT1 \cite{DAF-MN-Himsoon,DAF-DDF-QZ,DAF-General} and semi-OPT2 \cite{DAF-ITVT}, with simulation results.

The semi-OPT1 weights are written as \cite{DAF-Liu,DAF-DDF-QZ,DAF-General,DAF-MN-Himsoon},
\begin{equation}
\label{eq:w0w2_slow}
\begin{split}
w_0^{\sMRC}&=\frac{1}{2N_0},\\
w_2^{\sMRC}&=\frac{1}{2N_0(1+A^2\sigma_2^2)}.
\end{split}
\end{equation}
Also, the semi-OPT2 weights are written as \cite{DAF-ITVT}
\begin{equation}
\label{eq:w0w2_opt2}
\begin{split}
w_0^{\sOPT2}&=\frac{\alpha_0}{N_0[1+\alpha_0^2+(1-\alpha_0^2)\rho_0]},\\
w_2^{\sOPT2}&=\frac{\alpha}{N_0[(1+\alpha^2)(1+A^2\sigma_2^2)+(1-\alpha^2)A^2\rho_1 \sigma_2^2]}.
\end{split}
\end{equation}
Practically, the two sets of weights are the same for slow-fading channels with $\alpha_0=1,\alpha=1.$

The optimum power allocation values, reported in \cite{DAF-MN-Himsoon} are considered for three scenarios: symmetric channels, $[\sigma_0^2,\sigma_1^2,\sigma_2^2]=[1,1,1]$, non-symmetric channels with strong SR channel $[\sigma_0^2,\sigma_1^2,\sigma_2^2]=[1,10,1]$ and non-symmetric channels with strong RD channel $[\sigma_0^2,\sigma_1^2,\sigma_2^2]=[1,1,10]$.  The values are $q_{\opt}=0.66,0.54,0.80$ for symmetric, strong SR and strong RD channels, respectively.
To generate the channel coefficients, the simulation method in \cite{m2m-patel} is used. 

Based on the mobility of nodes, three fading cases, listed in Table~\ref{table:f0f1f2} are considered here.
In Case I, all nodes are assumed to be fixed or slowly moving such that $f_0=f_1=f_2=0.001.$ In case II, only Source is moving and Relay and Destination are static such that $f_0=.02, f_1=0.001,f_2=0.001$. Both SD and SR channels are time-varying and RD channel is static. In Case III, both Source and Relay are moving and Destination is fixed such that $f_0=0.05, f_1=0.03,f_2=0.001.$ All channels are time-varying in this case.
\begin{table}[!ht]
\begin{center}
\caption{Three fading cases.}
\label{table:f0f1f2}
  \begin{tabular}{ |c | c| c|c|c|  }
    \hline
				& $f_0$ & $f_1$ & $f_2$ & Channels status   \\ \hline\hline
{Case I }		& 0.001  & 0.001  & 0.001  & all are slow-fading \\ \hline
{Case II } 		& 0.02   & 0.001   &  0.001	& SD and SR are fast-fading \\ \hline
{Case III } 	& 0.05   & 0.02   &   0.001 & all are fast-fading \\
    \hline
  \end{tabular}
\end{center}
\end{table}

\begin{figure}[b!]
\psfrag {Lower Bound} [] [] [.7] {\qquad Lower Bound \cite{DAF-ITVT}}
\psfrag {P(dB)} [t][] [1]{$P/N_0$ (dB)}
\psfrag {BER} [] [] [1] {BER}
\centerline{\epsfig{figure={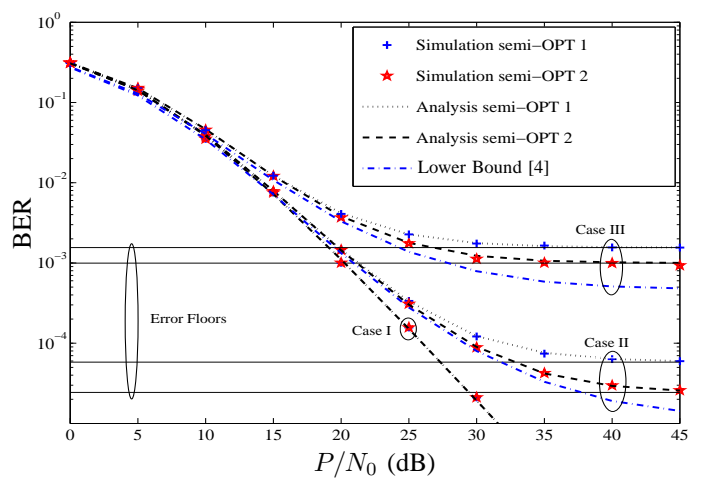},height=6cm,width=9cm}}
\caption{Theoretical and simulation BER of the D-AF system with the semi-OPT1 and semi-OPT2 weights in different fading rates and $[\sigma_0^2,\sigma_2^2,\sigma_3^2]=[1,1,1]$.}
\label{fig:m2_sig1_fds}
\end{figure}

\begin{figure}[b!]
\psfrag {Simulation semi-MRC} [] [] [.7] {$\quad$Simulation semi-OPT1}
\psfrag {Analysis semi-MRC} [] [] [.7] {$\quad$Analysis semi-OPT1}
\psfrag {Simulation EGC} [] [] [.7] {$\quad$Simulation EGC}
\psfrag {P(dB)} [t][] [1]{$P/N_0$ (dB)}
\psfrag {BER} [] [] [1] {BER}
\centerline{\epsfig{figure={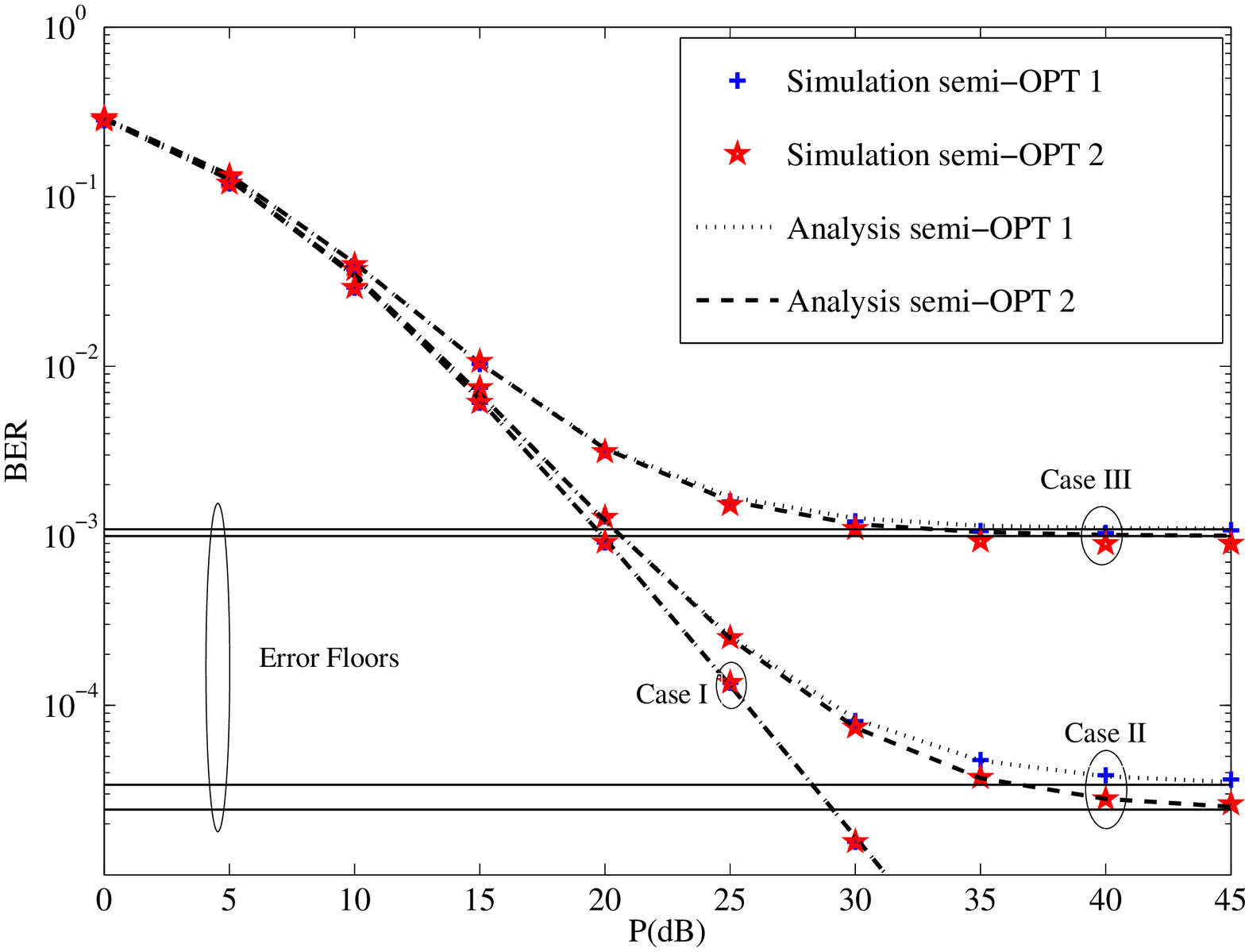},height=6cm,width=9cm}}
\caption{Theoretical and simulation BER of the D-AF system with the semi-OPT1 and semi-OPT2 weights in different fading rates and $[\sigma_0^2,\sigma_2^2,\sigma_3^2]=[1,10,1]$.}
\label{fig:m2_sig2_fds}
\end{figure}

\begin{figure}[b!]
\psfrag {Simulation semi-MRC} [] [] [.7] {$\quad$Simulation semi-OPT1}
\psfrag {Analysis semi-MRC} [] [] [.7] {$\quad$Analysis semi-OPT1}
\psfrag {Simulation EGC} [] [] [.7] {$\quad$Simulation EGC}
\psfrag {Analysis EGC} [] [] [.7] {$\quad$Analysis EGC}
\psfrag {P(dB)} [t][] [1]{$P/N_0$ (dB)}
\psfrag {BER} [] [] [1] {BER}
\centerline{\epsfig{figure={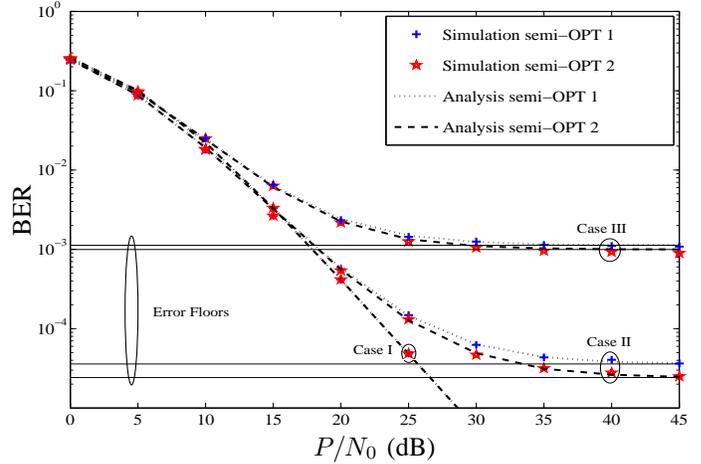},height=6cm,width=9cm}}
\caption{Theoretical and simulation BER of the D-AF system with the semi-OPT1 and semi-OPT2 weights in different fading rates and $[\sigma_0^2,\sigma_2^2,\sigma_3^2]=[1,1,10]$.}
\label{fig:m2_sig3_fds}
\end{figure}

The simulated and theoretical BER values using the semi-OPT1 and semi-OPT2 weights are computed for all cases and different channel variances and plotted versus $P/N_0$ in Figs.~\ref{fig:m2_sig1_fds}-~\ref{fig:m2_sig3_fds}. The theoretical values of the error floor for Cases II and III are computed and added to the figures. The error floor for Case I is very small and hence not plotted. For comparison purposes, the theoretical lower bound derived in \cite{DAF-ITVT} for symmetric channels is also added to Fig.~\ref{fig:m2_sig1_fds}.

From Figs.~\ref{fig:m2_sig1_fds}-\ref{fig:m2_sig3_fds}, it can be seen that the simulation and our analytical results in all fading and channel scenarios are tight to each other for both semi-OPT1 and semi-OPT2 weights (whereas the lower bound of \cite{DAF-ITVT} is relatively loose in Cases II and III of Fig.\ref{fig:m2_sig1_fds}). Specifically, in Case I of all figures (slow-fading channels), the desired diversity is achieved as expected. The combining weights of semi-OPT1 and semi-OPT2 are practically the same in this case and their results overlap. On the other hand, in Cases II and III (fast-fading channels) of all the figures the BER results gradually deviate from slow-fading results and reach to error floors as predicted.
As expected, in Cases II and III of the figures, the results of semi-OPT2 are slightly better than semi-OPT1 at high SNR region. The reason is that the semi-OPT2 weights take channel auto-correlation values into account so that they are closer to optimum weights.



\balance

\section{Conclusion}
\label{sec:con}
Differential amplify-and-forward relaying using linear combining with arbitrary fixed combining weights in time-varying Rayleigh fading channels was considered. The BER of the system employing DBPSK was derived and verified with simulation for two sets of combining weights. With the obtained BER expression, the effect of channel variation on the system performance was also studied. 

\bibliographystyle{IEEEbib}
\bibliography{ref/references}

\end{document}